# Data-driven Exploration of Pressure-Induced Superconductivity in $AgIn_5Se_8$

Ryo Matsumoto[a,d], Hiroshi Hara[a,d], Zhufeng Hou[b], Shintaro Adachi[a], Hiromi Tanaka[e], Sayaka Yamamoto[a,d,e], Yoshito Saito[a,d], Hiroyuki Takeya[a], Tetsuo Irifune[f], Kiyoyuki Terakura[c] and Yoshihiko Takano[a,d]

[a]*International Center for Materials Nanoarchitectonics (MANA), National Institute for Materials Science, 1-2-1 Sengen, Tsukuba, Ibaraki 305-0047, Japan*
[b]*Research and Services Division of Materials Data and Integrated System (MaDIS), National Institute for Materials Science, 1-2-1 Sengen, Tsukuba, Ibaraki 305-0047, Japan*
[c]*Center for Materials research by Information Integration (CMI$^2$), National Institute for Materials Science, 1-2-1 Sengen, Tsukuba, Ibaraki 305-0047, Japan*
[d]*University of Tsukuba, 1-1-1 Tennodai, Tsukuba, Ibaraki 305-8577, Japan*
[e]*National Institute of Technology, Yonago College, 4448 Hikona, Yonago, Tottori 683-8502, Japan*
[f]*Geodynamics Research Center, Ehime University, Matsuyama, Ehime 790-8577, Japan*

**Abstract**

Candidates compounds for new thermoelectric and superconducting materials, which have narrow band gap and flat bands near band edges, were exhaustively searched by a high-throughput first-principles calculation from an inorganic materials database named AtomWork. We focused on $AgIn_5Se_8$ which has high density of state near the Fermi level. $AgIn_5Se_8$ was successfully synthesized as single crystals using a melt and slow cooling method. The single-crystal X-ray diffraction analysis revealed the obtained crystal is high quality without deficiencies. The valence states in $AgIn_5Se_8$ were determined to be $Ag^{1+}$, $In^{3+}$ and $Se^{2-}$ in accordance with a formal charge by the core level X-ray photoelectron spectroscopy analysis. The electrical resistance was evaluated under high pressure using a diamond anvil cell with boron-doped diamond electrodes. Although the sample was insulator with a resistance of above 40 MΩ at ambient pressure, the resistance markedly decreased with increase of the pressure, and a pressure-induced superconducting transition was discovered at 3.4 K under 52.5 GPa. The transition temperature increased up to 3.7 K under further pressure of 74.0 GPa.

## 1. Introduction

The idea of using informatics techniques in conjunction with the data-driven approach based on high-throughput computation for functional materials design, namely materials informatics, recently has been carried out in practice and also has produced remarkable outcomes [1-5]. For example, a combination of deep learning-assisted artificial intelligence (AI) and density functional theory (DFT) calculations enabled researchers within 10 days to find five synthesizable and stable organic molecules with desired properties of absorption wavelength [6]. On the other hand, almost all superconducting materials have been discovered via the traditional carpet-bombing type experiments which were based on the experiences and inspirations of researchers.

We have performed the data-driven approach to explore new pressure-induced superconductors using a database and the DFT calculation [7,8]. In this particular screening, candidate compounds were selected by a specific band structure of "flat band" near a Fermi level [9]. If the flat band crosses the Fermi level, superconductivity would be realized due to high a density of state (DOS) [10-12]. For example, an existence of singularity of DOS, known as a van Hove singularity (vHs) is predicted near the Fermi level in compressed $H_3S$ with highest superconducting transition temperature ($T_c$) [11] and high-$T_c$ cuprates [13,14]. Indeed, new pressure-induced superconductivity was discovered in real compounds of $SnBi_2Se_4$ [7] and $PbBi_2Te_4$ [8] through the high-throughput screening.

One of the most interesting material is $AgIn_5Se_8$ in the screened-out candidates. There is no research considering this compound as a candidate of superconductor because of a relatively wide band gap [15]. $AgIn_5Se_8$ is $In_2Se_3$-based pseudo-binary alloy, belongs to the $A^I B^{III}_5 C^{VI}_8$-type semiconductor family which is focused on a candidate of solar cell materials [16]. Carrier-tuned $AgIn_5Se_8$ is also expected as a candidate for a superior thermoelectric material [15, 17]. Moreover, since $AgIn_5Se_8$ equips high DOS near valence band edges, it could be expected the appearance of superconductivity by band-structure engineering using high pressure technique.

To examine the appearance of superconductivity under high pressure, we have synthesized the sample of $AgIn_5Se_8$ in single crystals. The crystal structure, compositional ratio, and valence state of $AgIn_5Se_8$ single crystals were analyzed by the X-ray diffraction (XRD), energy dispersive X-ray spectrometry (EDX), and X-ray photoelectron spectroscopy (XPS), respectively. The electrical resistance of the obtained sample was evaluated under high pressure using a diamond anvil cell (DAC) with boron-doped diamond electrodes [18-21].

## 2. Electronic band structure under high pressures

Electronic band structures of the selected material $AgIn_5Se_8$ were calculated under various pressures up to 70 GPa. The details of our screening scheme in the high-throughput first-principles calculations were given in our previous paper [7,8]. Figure 1 shows the band structures and the total DOSs of $AgIn_5Se_8$ up to 70 GPa, obtained by the generalized gradient approximation. The valence band edges of $AgIn_5Se_8$ show a flat feature near the Fermi level. The band gap under ambient pressure is 0.24 eV which is wider than that of previous compounds of $SnBi_2Se_4$ and $PbBi_2Te_4$. Although the DOS at valence band edge increases with quite sharp peak like the van Hove singularity under 10 GPa, the band gap dose not decrease. The band gap gradually decreases with the increase of the pressure from 20 GPa, and completely closes at 50 GPa, indicating an insulator-to-metal transition. According to the insensitive response for the applied pressure, critical pressure for superconductivity in AgIn$_5$Se$_8$ would be higher than 10-20 GPa of SnBi$_2$Se$_4$ and PbBi$_2$Te$_4$. The $T_c$ would be enhanced under further pressure because the DOS near the Fermi level increases from 50 GPa to 70 GPa.

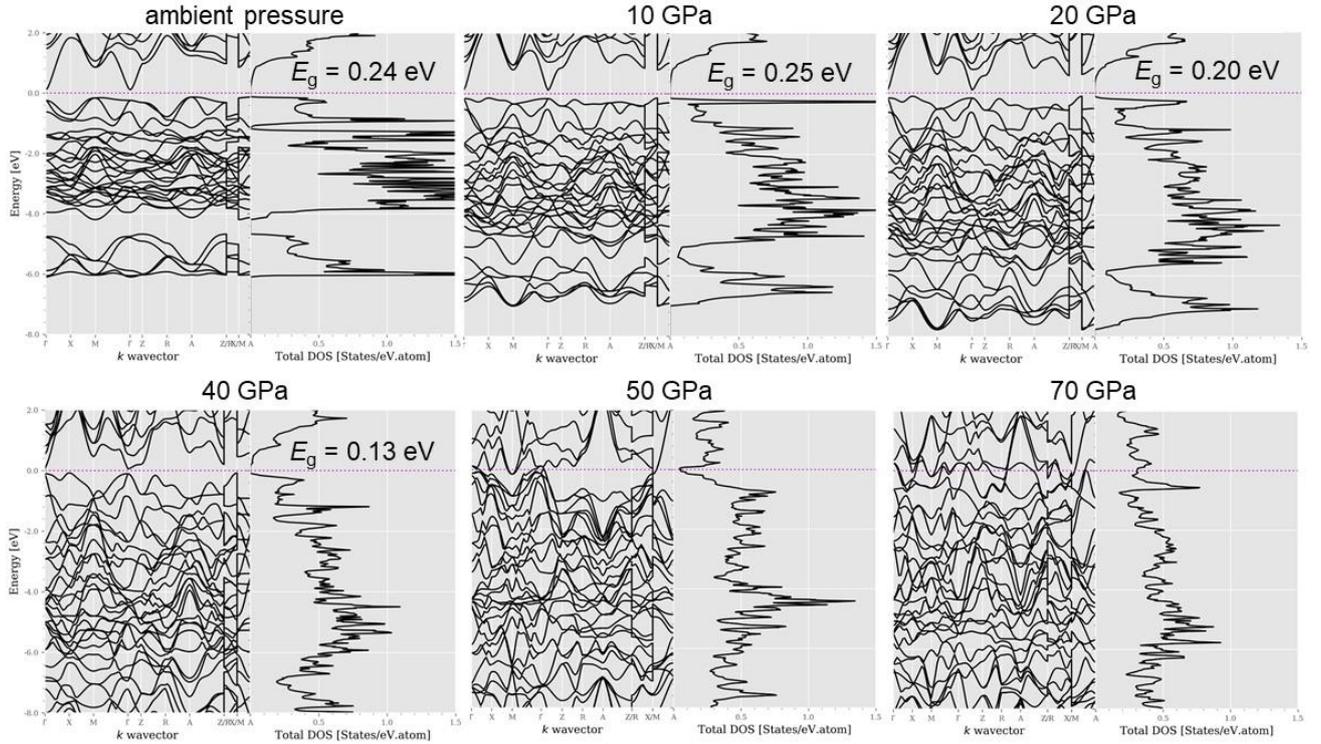

**Figure 1. Band structures and total density of states (DOS) of AgIn$_5$Se$_8$ up to 70 GPa, obtained by the generalized gradient approximation.**

## 3. Experimental procedures

3.1 Sample synthesis

Single crystals of AgIn$_5$Se$_8$ were grown by a melt and slow-cooling method. Starting materials of Ag powders (99.99%), In grains (99.98%) and Se chips (99.999%) were weighed in the stoichiometric composition of AgIn$_5$Se$_8$. The starting materials were loaded into an evacuated quartz tube and pre-reacted at 500ºC for 10 hours. The resultant tube was heated at 1000ºC for 20 hours and then slowly cooled to 600ºC in 20 hours, followed by the homogenizing anneal at 500ºC for 20 hours.

3.2 Characterization

The single crystal XRD measurement was carried out by use of the XtaLAB mini (Rigaku) with Mo-K$\alpha$ radiation ($\lambda$ = 0.71072 Å). The crystal structure was solved by the program SHELXT on the WinGX software and refined using the program ShelXle [22-24]. The crystal structure was depicted by the software VESTA [25]. The chemical composition was determined by the EDX, using the JSM-6010LA (JEOL).

The valence state was estimated by the core level XPS analysis using the AXIS-ULTRA DLD (Shimadzu/Kratos) with monochromatic Al K$\alpha$ X-ray radiation ($h\nu$ = 1486.6 eV), operating under a pressure of the order of $10^{-9}$ Torr. The samples were cleaved using scotch tape in a highly vacuumed pre-chamber of the order of $10^{-7}$ Torr. A binding energy scale was established by referencing the C 1$s$ value of an adventitious carbon. A background signal was subtracted by using the active Shirley method implemented in COMPRO software [26]. The photoelectron peaks were analyzed by a pseudo-Voigt functions peak fitting.

3.3 Transport property measurement

Electrical resistance measurements of AgIn$_5$Se$_8$ single crystal under high pressure were performed using an originally designed DAC with boron-doped diamond electrodes [18-21]. Figure 2 shows an optical image of the sample space of our DAC. The sample was placed at the center of a bottom nano-polycrystalline diamond anvil [27] where the boron-doped diamond electrodes were fabricated. The undoped diamond insulating layer separates the electrodes and a metal gasket. The details of the cell configuration were described in a literature [20]. Cubic boron nitride powders with a ruby manometer were used as a pressure-transmitting medium. The applied pressure values were estimated by a fluorescence from ruby powders [28] and a Raman spectrum from the culet of top diamond anvil [29] using the inVia Raman Microscope (RENISHAW).

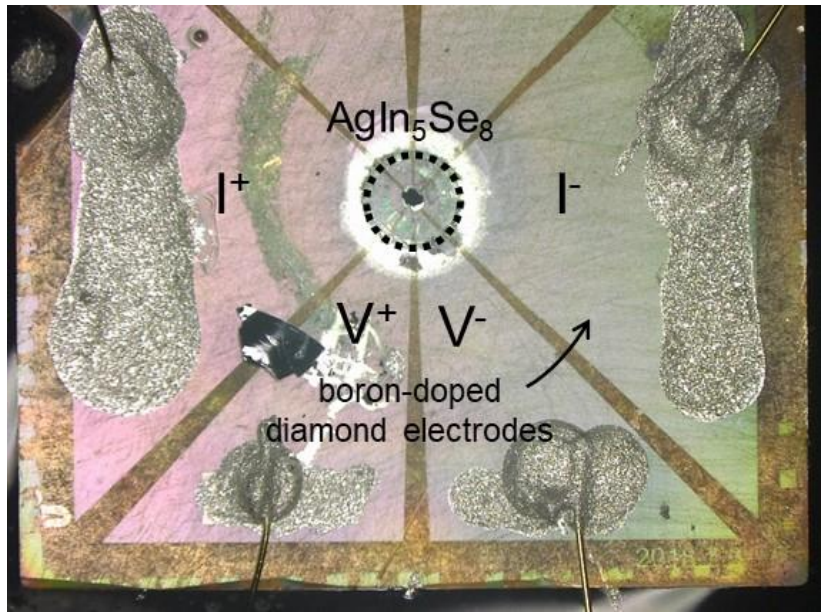

**Figure 2. Optical image of a bottom nano-polycrystalline diamond anvil with boron-doped diamond electrodes**.

**4. Results and discussion**
4.1 Structural analysis and composition

The single crystal structural analysis of AgIn$_5$Se$_8$ was successfully performed. The crystallographic data are given in supplemental tables SI-SIV. The final refinement with 22 variables using 1 restraint converged on the $R_1$ value of 6.77% and the $wR_2$ value of 15.14% for $I > 2\sigma(I)$ with the goodness of fit of 1.297. The refinement well converged when all the site occupancies were unity. There was no significant difference in $R_1$ values as the occupancy of a specific site was varied, indicating that the obtained AgIn$_5$Se$_8$ does not have any deficiencies at each site. The chemical composition of the obtained single crystal is normalized by the Ag element and then is estimated to be Ag:In:Se = 1:4.9(1):8.0(1) from the EDX analysis, which is consistent with the nominal composition of $AgIn_5Se_8$. This fact is different from the case of $SnBi_2Se_4$ which exhibits the deficiencies at Bi and Se sites [7]. The crystal structure was assigned as a tetragonal lattice (space group: *P-42m*) with lattice constants of $a = b = 5.7733(15)$ Å and $c = 11.584(4)$ Å. The ball-and stick model for the crystal structure of $AgIn_5Se_8$ is shown in Fig. 3. The In and Se atoms form two kinds of $InSe_4$ tetrahedrons: one is composed of In1, Se1 and Se2 atoms (Tet. A), and the other consists of In2 and Se2 atoms (Tet. B). The Tet. A and the Tet. B form each $InSe_4$ tetrahedron layer, and the crystal structure of $AgIn_5Se_8$ is upper and lower symmetric around the central Ag atom. The sum of the central angles in the Tet. A of 432.3º is near to that in regular tetrahedron (423.16º) using the bond angles in Table SIV, where a central angle is defined as an angle between a central atom and apical atoms of tetrahedron. While, the sum of the central angles in the Tet. B of 447º is far from the value of 423.16º, indicating that the Tet. B is more distorted than the Tet. A.

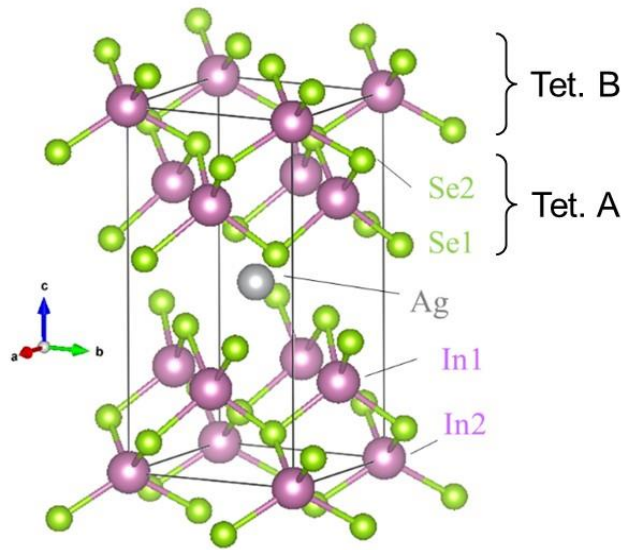

**Figure 3. A schematic image of the crystal structure of $AgIn_5Se_8$.**

4.2 Valence state

The valence states of Ag, In and Se in $AgIn_5Se_8$ were investigated by the XPS measurement. Figure 4(a) shows Ag 3d core-level spectrum of $AgIn_5Se_8$. There are two main peaks at 367.7 eV and 373.7 eV corresponding to Ag $3d_{5/2}$ and $3d_{3/2}$ with the valence state of $Ag^{1+}$ [30]. Figure 4(b) displays In 3d core-level spectrum. A perfect fit to two peaks located at 444.7 and 452.3 eV are clearly observed, which can be attributed to the characteristic spin-orbit split $3d^{5/2}$ and $3d^{3/2}$, respectively [31]. This result indicates that the indium valency is mainly 3+ in the crystal [32]. The peak of Se is more complex, as seen in fig. 4(c), because the spin-orbit-splitting of 0.86 eV in the Se 3d core levels is narrower [33]. The Se $3d^{5/2}$ and $3d^{3/2}$ peaks are located at 54.0 eV and 54.8 eV, respectively. From this fitting, it can be concluded that the valence state of Se is mainly 2-. These $Ag^{1+}$, $In^{3+}$ and $Se^{2-}$ are consistent with a formal charge valence of $AgIn_5Se_8$.

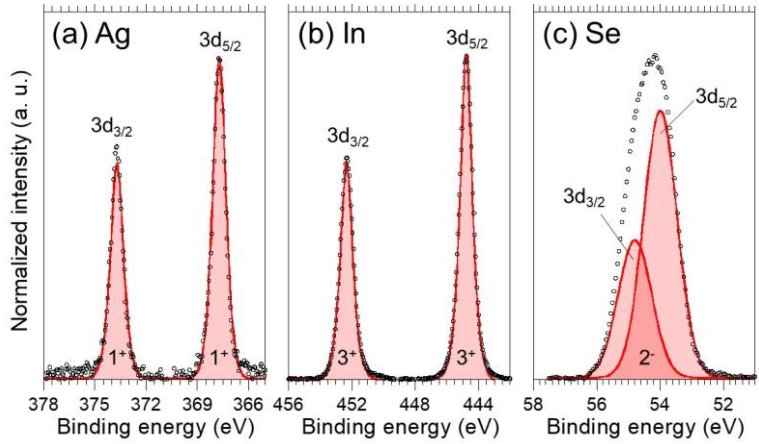

**Figure 4. High-resolution XPS spectra of (a) Ag 3d, (b) In 3d, and (c) Se 3d core levels in AgIn$_5$Se$_8$ single crystal.**

4.3 Electrical resistance measurement under high pressure

The electrical resistance of AgIn$_5$Se$_8$ was measured under high pressure to investigate the pressure-induced superconductivity. The sample at ambient pressure behaved as insulator with quite high resistance above 40 MΩ. The resistance was decreased less than 100 Ω at 7.0 GPa, and then we evaluated the temperature dependence of resistance under various pressures up to 24.8 GPa, as shown in Fig. 5(a) with log scale. The resistance decreased more than two orders of magnitude with increase of applied pressures, and the slope d$R$/d$T$ became almost flat from 24.8 GPa indicating a metallization. This feature of band gap closing is good agreement with the theoretical estimation of band gap closing as shown in Fig.1. Figure 5(b) shows the temperature dependence of resistance from 38.9 GPa to the highest pressure of 85.2 GPa. The sample under 52.5 GPa exhibited a sudden drop of resistance from 3.4 K corresponding to superconductivity. The pressure-induced insulator to metal transition and the superconductivity in AgIn$_5$Se$_8$ were successfully discovered based on our data-driven strategy. The superconductivity remained up to 85.2 GPa, and the diamond anvil was broken above further pressure, unfortunately.

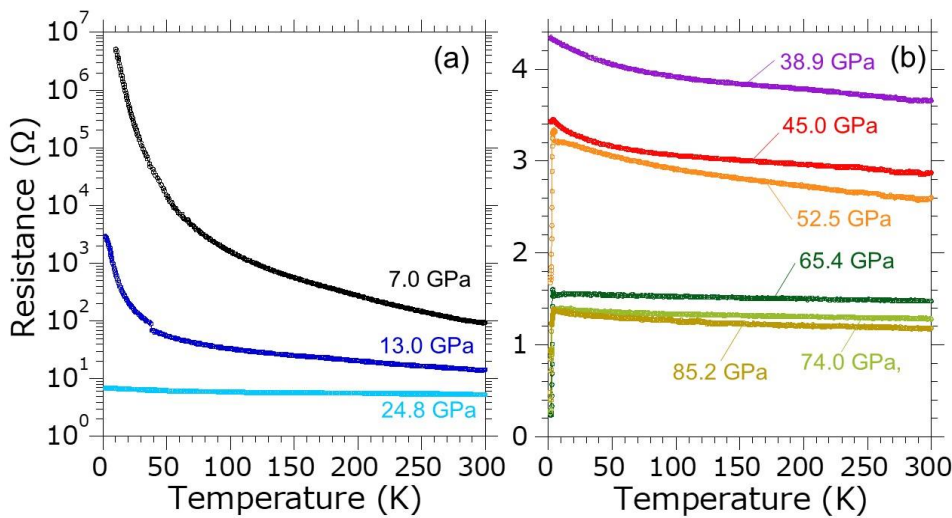

**Figure 5. Temperature dependence of resistance in AgIn$_5$Se$_8$ under various pressures of (a) 7.0 – 24.8 GPa in log scale, (b) 38.9 - 85.2 GPa in linear scale.**

Figure 6 (a) shows an enlargement of the temperature dependence of resistance around the superconducting transitions. The $T_c$ of AgIn$_5$Se$_8$ was almost constant around 3.4 K from 52.5 GPa to 65.4 GPa. The behavior was drastically changed above 74.0 GPa with a multi-step superconducting transition and an increase of $T_c$, and the $T_c$ slightly decreased at 85.2 GPa. To clear the change of behavior between 65.4 GPa and 74.0 GPa, we picked up the data from each pressure region as shown in Fig. 6 (b). The resistance under 65.4 GPa was increased around 3.5 K just before the onset of superconducting transition, which feature is known for the inhomogeneity of superconducting region [34]. To confirm that the hump is originated from the superconductivity, we measured the temperature dependence of resistance under magnetic field as shown in Fig. 7. The hump at just before the drop of resistance under 65.4 GPa was suppressed by applying magnetic field, therefore, this would be sign for appearance of higher $T_c$ phase. Instead of the suppression of the hump, the three steps of superconducting transition appeared at 74 GPa, which labeled by lower $T_c$, middle $T_c$, and higher $T_c$ of 2.3 K, 2.9 K, and 3.8 K, respectively. There is a possibility of crystal structural phase transition of this compound above 50 GPa by our calculations for structural stability under various pressures. Our DFT calculations showed that the lattice constant of $c$ axis was shrunk and $a(b)$ axis was expanded when the applied pressure was around 50 GPa. The suddenly appeared three steps superconducting transition would be originated from the intrinsically distorted crystal structure and structural phase transition of AgIn$_5$Se$_8$. The theoretical exploration of possible structural phase transition of AgIn$_5$Se$_8$ is undertaken and we will discuss the results in a future work. All $T_c$ values under 74.0 GPa were gradually suppressed by an increase of magnetic field, and completely disappeared above 2 K under 4 T. Although the lower and middle $T_c$ was decreased under the highest pressure of 85.4 GPa, the higher $T_c$ stayed around 3.8 K.

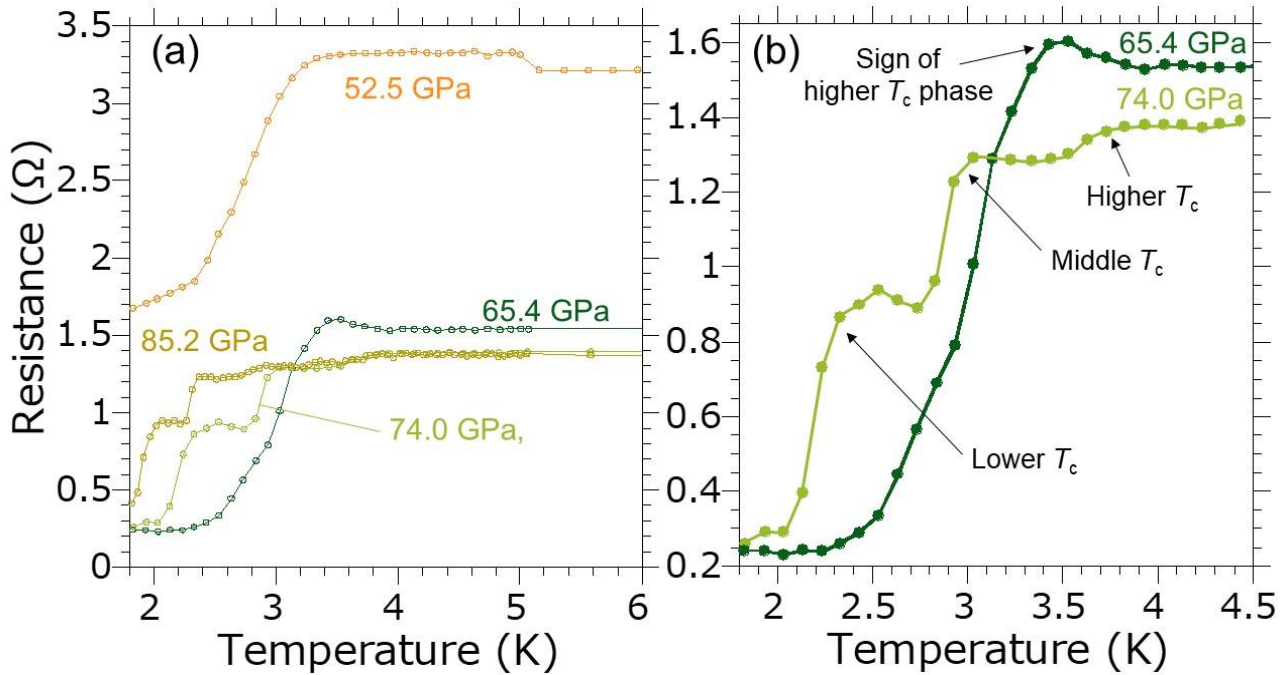

**Figure 6. (a) Temperature dependence of resistance around superconducting transitions in AgIn$_5$Se$_8$ from 52.5 GPa to 85.2 GPa. (b) Picked up data under 65.4 GPa and 74.0 GPa.**

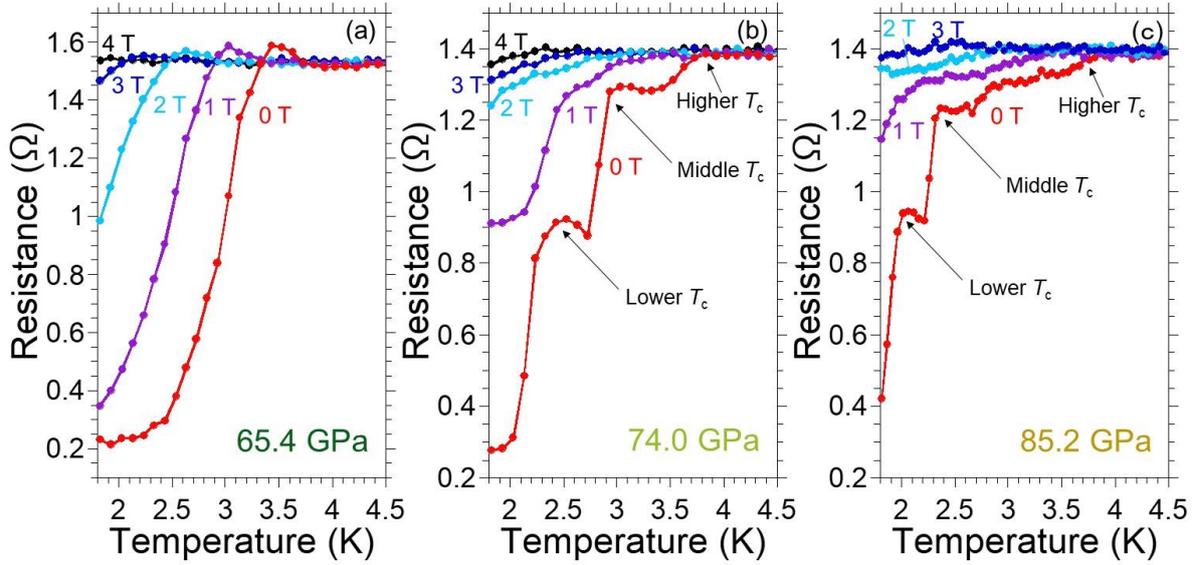

**Figure 7.** Temperature dependence of resistance of $AgIn_5Se_8$ in magnetic fields under (a) 65.4 GPa, (b) 74.0 GPa and (c) 85.2 GPa.

Figure 8 shows a pressure-dependent phase diagram of $AgIn_5Se_8$ single crystal. The sample at ambient pressure was insulator with quite high resistance above 40 MΩ. The resistance markedly decreased under 10-20 GPa, caused by the band gap closing. The temperature dependence of resistance became metallic tendency from around 30 GPa, and the pressure-induced superconductivity appeared above 50.2 GPa. The $T_c$ of $AgIn_5Se_8$ increased from 74.0 GPa in accordance with the enhancement of DOS near the Fermi level. The maximum $T_c$ of 3.8 K was observed under 74.0 and 85.2 GPa. When the applied pressure increased further, the diamond anvil was broken.

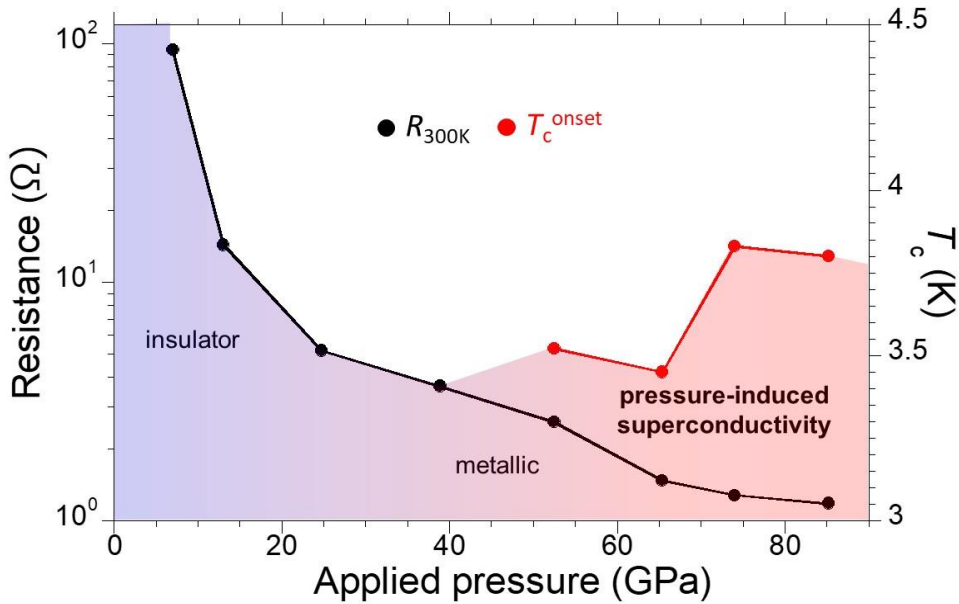

**Figure 8.** Pressure-dependent phase diagram of $AgIn_5Se_8$.

## 5. Conclusion

We focused on $AgIn_5Se_8$ from the viewpoint of the pressure-induced high DOS near the Fermi level, compared with the other candidates of SnBi$_2$Se$_4$ and PbBi$_2$Te$_4$ which was also chosen by the data-driven approach. The single crystal growth of AgIn$_5$Se$_8$ was confirmed by single crystal XRD and EDX analysis. The XPS analysis revealed that the valence state of all elements was consistent with the nominal value. The synthesized AgIn$_5$Se$_8$ showed insulator to metal transition at 24.8 GPa and superconducting transition at 52.5 GPa by the high pressure resistance measurement using DAC with boron-doped diamond electrodes in accordance with our scenario. The highest $T_c$ was observed at 3.8 K under 74.0 and 85.2 GPa. The pressure-induced metallization and superconductivity were successfully observed in AgIn$_5$Se$_8$. According to the multi-steps of $T_c$ above 74 GPa, it is necessary to perform the further studies for the crystal structural analysis under higher pressure.

**Acknowledgment**

This work was partly supported by JST CREST Grant No. JPMJCR16Q6, JST-Mirai Program Grant Number JPMJMI17A2, JSPS KAKENHI Grant Number JP17J05926, and the "Materials research by Information Integration" Initiative (MI$^2$I) project of the Support Program for Starting Up Innovation Hub from JST. A part of the fabrication process of diamond electrodes was supported by NIMS Nanofabrication Platform in Nanotechnology Platform Project sponsored by the Ministry of Education, Culture, Sports, Science and Technology (MEXT), Japan. The part of the high pressure experiments was supported by the Visiting Researcher's Program of Geodynamics Research Center, Ehime University. The computation in this study was performed on Numerical Materials Simulator at NIMS.

**References**
[1] T. Ishikawa, T. Oda, N. Suzuki, and K. Shimizu, High Pressure Res. **35**, 37 (2015).
[2] T. Ishikawa, A. Nakanishi, K. Shimizu, H. Katayama-Yoshida, T. Oda, and N. Suzuki, Sci. Rep. **6**, 23160 (2016).
[3] S. Kiyohara, H. Oda, T. Miyata, and T. Mizoguchi, Sci. Adv. **2**, e1600746 (2016).
[4] A. Seko, A. Togo, H. Hayashi, K. Tsuda, L. Chaput, and I. Tanaka, Phys. Rev. Lett. **115**, 205901 (2015).
[5] T. Inoshita, S. Jeong, N. Hamada, and H. Hosono, Phys. Rev. X **4**, 031023 (2014).
[6] M. Sumita, X. Yang, S. Ishihara, R. Tamura, and K. Tsuda, ASC Cent. Sci **4**, 1126-1133 (2018).
[7] R. Matsumoto, Z. Hou, H. Hara, S. Adachi, H. Takeya, T. Irifune, K. Terakura, and Y. Takano, Appl. Phys. Express **11**, 093101 (2018).
[8] R. Matsumoto, Z. Hou, M. Nagao, S. Adachi, H. Hara, H. Tanaka, K. Nakamura, R. Murakami, S. Yamamoto, H. Takeya, T. Irifune, K. Terakura, and Y. Takano, Sci. Technol. Adv. Mater. **19**, 909-916 (2018).
[9] K. Kuroki and R. Arita, J. Phys. Soc. Jpn. **76**, 083707 (2007).
[10] K. S. Fries and S. Steinberg, Chem. Mater. **30**, 2251 (2018).
[11] W. Sano, T. Koretsune, T. Tadano, R. Akashi, and R. Arita, Phys. Rev. B **93**, 094525 (2016).
[12] Y. Ge, F. Zhang, and Y. Yao, Phys. Rev. **B** 93, 224513 (2016).
[13] A. A. Abrikosov, J. C. Campuzano, K. Gofron, Phys. Status Solidi C **214**, 73 (1993).

[14] D. M. Newns, C. C. Tsuei, P. C. Pattniak, Phys Rev. B **52**, 13611 (1995).

[15] J. L. Cui, Y. Y. Li, Y. Deng, Q. S. Meng, Y. L. Gao, H. Zhou, and Y. P. Li, Intermetallics **31**, 217-224 (2012).

[16] E. Ghorbani, J. Kiss, H. Mirhosseini, G. Roma, M. Schmidt, J. Windeln, T. D. Kühne, and C. Felser, J. Phys. Chem. C **119**, 25197-25203 (2015).

[17] S. Welzmiller, F. Hennersdorf, R. Schlegel, A. Fitch, G. Wagner, and O. Oeckler, Inorg. Chem **54**, 5745-5756 (2015).

[18] R. Matsumoto, Y. Sasama, M. Fujioka, T. Irifune, M. Tanaka, T. Yamaguchi, H. Takeya, and Y. Takano, Rev. Sci. Instrum. **87**, 076103 (2016).

[19] R. Matsumoto, T. Irifune, M. Tanaka, H. Takeya, and Y. Takano, Jpn. J. Appl. Phys. **56**, 05FC01 (2017).

[20] R. Matsumoto, A. Yamashita, H. Hara, T. Irifune, S. Adachi, H. Takeya, and Y. Takano, Appl. Phys. Express **11**, 053101 (2018).

[21] R. Matsumoto, H. Hara, H. Tanaka, K. Nakamura, N. Kataoka, S. Yamamoto, T. Irifune, A. Yamashita, S. Adachi, H. Takeya, and Y. Takano, J. Phys. Soc. Jpn. (accepted).

[22] G. M. Sheldrick, Acta Crystallogr., Sect. A **71**, 3 (2015).

[23] L. J. Farrugia, J. Appl. Cryst. **45**, 849 (2012).

[24] C. B. Hübschle, G. M. Sheldrick and B. Dittrich, J. Appl. Cryst. **44**, 1281 (2011).

[25] K. Momma and F. Izumi, J. Appl. Cryst. **44**, 1272 (2011).

[26] R. Matsumoto, Y. Nishizawa, N. Kataoka, H. Tanaka, H. Yoshikawa, S. Tanuma, and K. Yoshihara, J. Electron Spectrosc. Relat. Phenom. **207**, 55 (2016).

[27] T. Irifune, A. Kurio, S. Sakamoto, T. Inoue, and H. Sumiya, Nature **421**, 599 (2003).

[28] G. J. Piermarini, S. Block, J. D. Barnett, and R. A. Forman, J. Appl. Phys. **46**, 2774 (1975).

[29] Y. Akahama and H. Kawamura, J. Appl. Phys. **96**, 3748 (2004).

[30] S. Aravindan, V. Rajendran, and N. Rajendran, Phase Transitions **85**, 630-649 (2012).

[31] Y. S. Kang, C. Y. Kim, M. H. Cho, K. B. Chung, C. H. An, H. Kim, H. J. Lee, C. S. Kim, and T. G. Lee, Appl. Phys. Lett. **97**, 172108 (2010).

[32] L. Yang, T. Wang, Y. Zou, and H. L. Lu, Nanoscale Res. Lett. **12**, 339 (2017).

[33] B. Zhan, S. Butt, Y. Liu, J. L. Lan, C. W. Nan, and Y. H. Lin, J. Electroceramics **34**, 175-179 (2015).

[34] G. Zhang, T. Samuely, J. Kačmarčík, E. A. Ekimov, J. Li, J. Vanacken, P. Szabó, J. Huang, P. J. Pereira, D. Cerbu, and V. V. Moshchalkov, Phys. Rev. Applied **6**, 064011 (2016).

**Supplemental materials**

**Table SI. Crystallographic data on the AgIn$_5$Se$_8$.**

| | |
|---|---|
| Structural formula | AgIn$_5$Se$_8$ |
| Formula weight (g/mol) | 1313.65 |
| Crystal dimensions (mm) | 0.2 × 0.28 × 0.06 |
| Crystal system | Tetragonal |
| Space group | *P-42m* (No. 111) |
| *a* (Å) | 5.7733(15) |
| *b* (Å) | 5.7733(15) |
| *c* (Å) | 11.584(4) |
| *V* (Å$^3$) | 386.1(2) |
| *Z* | 1 |
| $d_{cal}$ (g/cm$^3$) | 5.650 |
| Temperature (K) | 293(2) |
| *λ* (Å) | 0.71073 (MoK*α*) |
| *μ* (mm$^{-1}$) | 27.399 |
| *θ*$_{max}$ (º) | 27.42 |
| Index ranges | −7<*h*<7, −7<*k*<7, −14<*l*<14 |
| Total reflections | 1323 |
| Unique reflections | 249 |
| No. of variables | 22 |
| $R_{int}$ for all reflections | 0.1312 |
| $R_1$ / $wR_2$ [$I > 2\sigma(I)$] | 0.0677 / 0.1514 |
| $R_1$ / $wR_2$ (all data) | 0.0678 / 0.1515 |
| Goodness of fit | 1.297 |
| Max./Min. residual density (e$^-$/Å$^3$) | 13.617 / −3.436 |

**Table SII. Atomic coordinates and site occupancies of AgIn$_5$Se$_8$.**

| Atom | Wyck. | S.O.F. | *x* | *y* | *z* |
|---|---|---|---|---|---|
| Ag | 1*b* | 1 | 1/2 | 1/2 | 1/2 |
| In1 | 4*m* | 1 | 0 | 1/2 | 0.249(2) |
| In2 | 1*a* | 1 | 0 | 0 | 0 |
| Se1 | 4*n* | 1 | 0.225(3) | 0.225(3) | 0.6097(17) |
| Se2 | 4*n* | 1 | 0.273(2) | 0.273(2) | 0.1155(16) |

**Table SIII. Atomic displacement parameters (Å$^2$) of AgIn$_5$Se$_8$.**

| Atom | $U_{11}$ | $U_{22}$ | $U_{33}$ | $U_{23}$ | $U_{13}$ | $U_{12}$ | $U_{eq}$ |
|---|---|---|---|---|---|---|---|
| Ag | 0.040(12) | 0.040(12) | 0.007(16) | 0 | 0 | 0 | 0.029(10) |
| In1 | 0.013(5) | 0.029(7) | 0.0099(15) | 0 | 0 | 0.013(5) | 0.0173(12) |
| In2 | 0.008(6) | 0.008(6) | 0.011(14) | 0 | 0 | 0 | 0.009(6) |
| Se1 | 0.038(8) | 0.038(8) | 0.004(3) | −0.007(3) | −0.007(3) | 0.003(8) | 0.027(5) |
| Se2 | 0.006(4) | 0.006(4) | 0.008 | −0.005(2) | −0.005(2) | 0.002(5) | 0.006(3) |

Note: The $U_{33}$ of Se2 is fixed at the appropriate value to stabilize the refinement. $U_{eq}$ is defined as one-third of a trace of an orthogonalized $U$ tensor.

**Table SIV. Selected bond lengths (Å) and angles (º) of AgIn$_5$Se$_8$.**

| Bond length (Å) | | | |
|---|---|---|---|
| Ag–Se1 | 2.580(18) | In1–Se2 | 2.57(2) |
| In1–Se1 | 2.62(3) | In2–Se2 | 2.600(14) |
| Bond angle (º) | | | |
| Se1–In1–Se1 | 102.8(4) | Se2–In2–Se2 | 105.36(19) |
| Se1–In1–Se2 | 111.8(2) | | 105.4(3) |
| | 112.4(4) | | 118.1(3) |